# ENHANCING CT IMAGE SYNTHESIS FROM MULTI-MODAL MRI DATA BASED ON A MULTI-TASK NEURAL NETWORK FRAMEWORK

*Zhuoyao Xin[1]   Christopher Wu[2]   Dong Liu[3]   Chunming Gu[1,4,5]   Jia Guo[2*]   Jun Hua[1,4*]*

[1] F.M. Kirby Research Center for Functional Brain Imaging, Kennedy Krieger Institute, Baltimore, Maryland, United States; [2] Department of Biomedical Engineering, Columbia University, USA; [3] Department of Neuroscience, Columbia University, USA; [4] Neurosection, Division of MRI Research, Russell H. Morgan Department of Radiology and Radiological Science, Johns Hopkins University School of Medicine, Baltimore, Maryland, United States; [5] Department of Biomedical Engineering, Johns Hopkins University, Baltimore, Maryland, United States

**ABSTRACT**

Image segmentation, real-value prediction, and cross-modal translation are critical challenges in medical imaging. In this study, we propose a versatile multi-task neural network framework, based on an enhanced Transformer U-Net architecture, capable of simultaneously, selectively, and adaptively addressing these medical image tasks. Validation is performed on a public repository of human brain MR and CT images. We decompose the traditional problem of synthesizing CT images into distinct subtasks, which include skull segmentation, Hounsfield unit (HU) value prediction, and image sequential reconstruction. To enhance the framework's versatility in handling multi-modal data, we expand the model with multiple image channels. Comparisons between synthesized CT images derived from T1-weighted and T2-Flair images were conducted, evaluating the model's capability to integrate multi-modal information from both morphological and pixel value perspectives.

*Index Terms*— T1, FLAIR, synthetic CT, Multimodal, Multi-task, Transformer U-Net

## 1. INTRODUCTION

Synthetic CT is a useful technique to generate CT images from MRI. The application of synthetic CT image has garnered significant interest owing to their broad clinical utility, which includes like aberration correction in focused ultrasound (FUS) therapy [1, 2] and PET/MRI-CT attenuation correction [3, 4], both of which require simultaneous acquisition of multimodal image information. It leverages complementary features of the two modalities, namely CT's superior ability in outlining bone tissue and MRI's exceptional precision in delineating soft tissue [5, 6], which help to minimize costs, acquisition time, ionizing radiation exposure, while circumventing registration errors[7-9]. Notably, existing synthetic CT methods using generator-based methods or Generative Adversarial Networks (GANs) [10, 11] primarily focused on pixel-to-pixel predictions. These studies and the objective were to minimize the differences in pixel values across images.

Our approach dissects the MRI-to-CT transformation task into a dual pipeline: skull segmentation and the prediction of Hounsfield Unit (HU) pixels within the target region. In order to convert MR images into synthetic CT images, we introduced a versatile processing framework based on the Transformer U-Net. As the part of training method, the framework involves 3D patch-based strategy before sequential image reconstruction. It can effectively adapt to a wide array of imaging tasks, ranging from image segmentation to real-value prediction. Moreover, it is well-suited for cross-modal translation or reconstruction that necessitates comprehensive image structure and details.

The prior work often employed 2D [19] or pseudo-3D[10] training and evaluation to circumvent the challenges of large 3D image sizes, leading to the loss of continuity between slices and biased evaluations. To address these issues, we have incorporated 3D patch extraction and image sequential restoration into our framework. This allows to extract patches of various sizes from larger images, preserving 3D structural continuity and enhancing model generalizability, freeing training from being bound by the original size of input images. Finally, most existing methods exploit single modality MRI sequences such as T1, T2, and FLAIR [12-15] for CT synthesis, leveraging their inclusion in the widespread protocols of MRI and reduced technical prerequisites, as well as ultra-short (UTE) or zero-time echo (ZTE) [16-18] which enhanced signals in the vicinity of bone tissue, to achieve higher-quality CT synthesis. Yet, synthetic-CT approaches that leverage multimodal information are still limited in these endeavors. Thus, we propose the efficacy of integration of such information using the generation of CT images from dual-modal T1w and T2w-FLAIR data of the human brain as an example. To gauge the model's performance across both sub-tasks and the primary transformation task, we utilize a diverse set of metrics, considering differences at both the morphological and pixel levels.

## 2. METHODS

### 2.1. Data preparation

The publicly available CERMEP-IDB-MRXFDG dataset [20], acquired in 37 healthy adult subjects (age 38.11 ± 11.36 years; range 23 – 65, 54% women). The dataset includes matched brain images, specifically T1w Magnetization-Prepared Rapid Gradient-Echo (MPRAGE), T2 Fluid Attenuated Inversion Recovery (FLAIR) MRI (Siemens 1.5T scanner, voxel size of 1.2 x 1.2 x 1.2 mm³), and CT (Biograph mCT64) for each subject. Dataset was randomly split into training, validation, and test set in the ratio of 8:1:1. The MR and CT images were subjected to a simple normalization process to scale their values within the range of 0 to 1. Specifically, CT intensity $I_{CT\ Positive} = I_{CT\ Positive}/3071$. The skull labels were generated from the CT images using a thresholding method, removing foreground and background information.

### 2.2. Processing Framework

The overarching framework of our approach is illustrated in Figure 1. For the task of MRI to CT skull transformation, we employed two independent pipelines to address the subtasks of mask segmentation and region pixel value prediction. Paired MRI and CT images serve as the input and ground truth into the framework and are split into 3D patches based on their spatial requirements. From the CT images, two corresponding real skull patches and mask patches are generated. The skull and mask patches served as labels for the pixel prediction pipeline and the segmentation pipeline, respectively. The mask patches are also simultaneously fed into the pixel prediction pipeline to provide additional attention to the model's predictions within specific regions.

In our task segmentation approach, the original global pixel-to-pixel prediction is no longer necessary. We employed geometric morphological dilation algorithms to appropriately expand the masks, specifying the regions of interest when calculating mean squared error (MSE). This significantly reduced the computation scope. For instance, from 128x128x128 patches, the region calculated with a dilated mask spanning two voxels can be reduced to an average of approximately $3\times10^5$ voxel units, requiring only one fifteenth of the original computations. On the other hand, the morphological segmentation results necessitate calculations over the entire patch region using the Dice plus Binary Cross-entropy (BCE) loss. This is because the segmentation task requires the identification of skull details from the global structure of the MRI. Therefore, the overall pipeline's loss can be calculated as follows:

$$Loss = (1-\lambda)Dice_{loss\ Global} + \lambda BCE_{loss\ Global} + \frac{1}{d^2}\sum_{i=1}^{d}\sum_{j=1}^{d} MSE_{Local[i,j]} \quad (1)$$

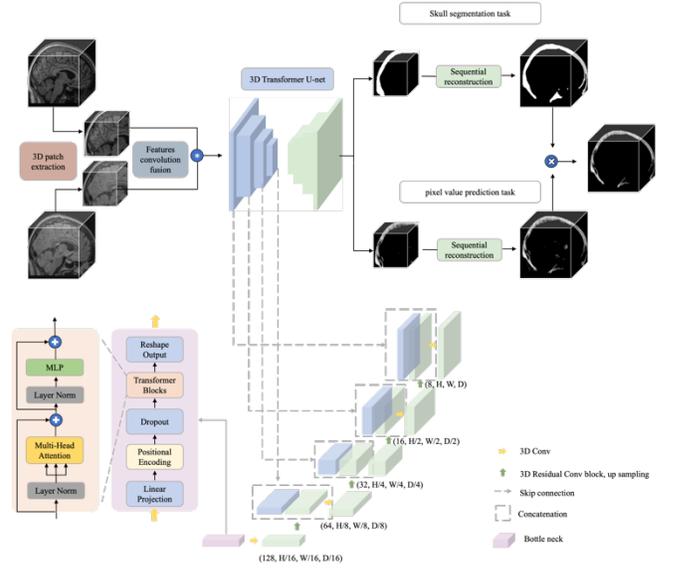

**Fig. 1.** Proposed 3D Transformer U-net based multi-task processing framework.

$\lambda$ controls the weight between the overall structural BCE and the detailed dice loss, while d denoted the number of geometric dilations. i and j represented the voxel's position in the patch. The model adopted a structure based on the enhanced Transformer U-Net [21], incorporating position encoding and multi-head attention mechanisms. In the encoder, features from different channels of the multimodal 3D images were combined through convolution for joint information prediction. The patch extraction positions were controlled by a switch, with the center of the patch falling within a specific region. The selection range for the center position was within an area that constitutes half the patch size on either side of the three-dimensional dimensions of the image (H, W, D). Therefore, theoretically, the total number of patches that can be generated from a single subject's image can be calculated as follow:

$$N_{Patch} = (1 + \frac{H - P_H}{S_h}) \times (1 + \frac{W - P_W}{S_w}) \times (1 + \frac{D - P_D}{S_d}) \quad (2)$$

Where P represents the patch size, and $S_{direction}$ is the step size each direction, which determines the overlap between patches. When S is greater than the side length of the patch in that direction, there is no overlap between the patches. The embedding of 3D patch extraction significantly expands the real dataset to a cubic level. This expansion enables the model to train without being limited by the size of the original dataset, thus avoiding the need for additional data augmentation steps. Even with only a few subjects' datasets, 3D data training is feasible because the framework has transformed the original global image task into a local structural learning task. This allows the model to focus on specific image structures. After the pipeline, the CT synthesis patches are sequentially filled back into the large image. For overlapping regions, we calculate the average value to enhance the stability of the overall prediction results. In the sequential image restoration, our image size is standardized to 207x243x226, with a patch size of 128 cubic, a step size of 6 in each direction, and a total of 5670 patch samples for one subject.

In this experiment, we employed a strategy of randomly selecting patch locations and sampling 100 patches from each subject's images in both training and validation sets. The total training set consisted of 3000 samples, and the validation set had 300 samples. To focus on information related to the bone structure, 100% of the patch samples used for pixel prediction were generated from the centers of the skull. For the segmentation task, we sampled 80% of the skull center points to generate patches, and the remaining 20% of the patch generation included areas near the skull and within the brain.

## 3. RESULT

**Fig.2** demonstrates the results of sub-tasks obtained using the multi-task processing framework. It includes a synthetic CT mask based on skull segmentation and a post-processed synthetic CT image derived from pixel-to-pixel predictions. In the top row, results are shown when using only T1 images as input to generate synthetic CT.

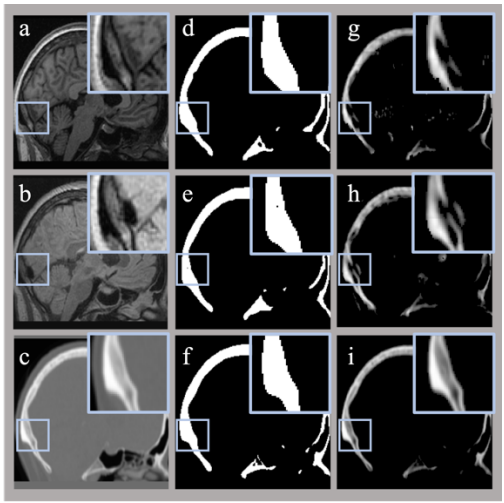

**Fig. 2.** Visualization of subtask results along the sagittal plane. (a-c) original T1w MRI, FLAIR MRI, and CT images, respectively. (d-f) Segmentation results obtained using the single T1w modality and the T1w+FLAIR modality, and the CT skull mask, respectively. (g-i) Pixel value prediction results using the single T1w modality and the T1w+FLAIR modality, and the ground truth CT images, respectively.

In the middle row, results are presented for CT image prediction based on both T1 and FLAIR modalities. **Fig.3** showcases a comprehensive synthetic CT result that combines both mask generation and pixel value prediction, with the local image details. It can be observed from the **Fig2**(g,h) that, even after geometrical mask thresholding, the pixel prediction sub-task of CT still exhibits residual errors in both foreground and background regions. This emphasizes the requirements for post-processing when segmentation masks are not used. However, for the synthetic CT generated by the integrated dual-pipeline approach, no additional post-processing was required.

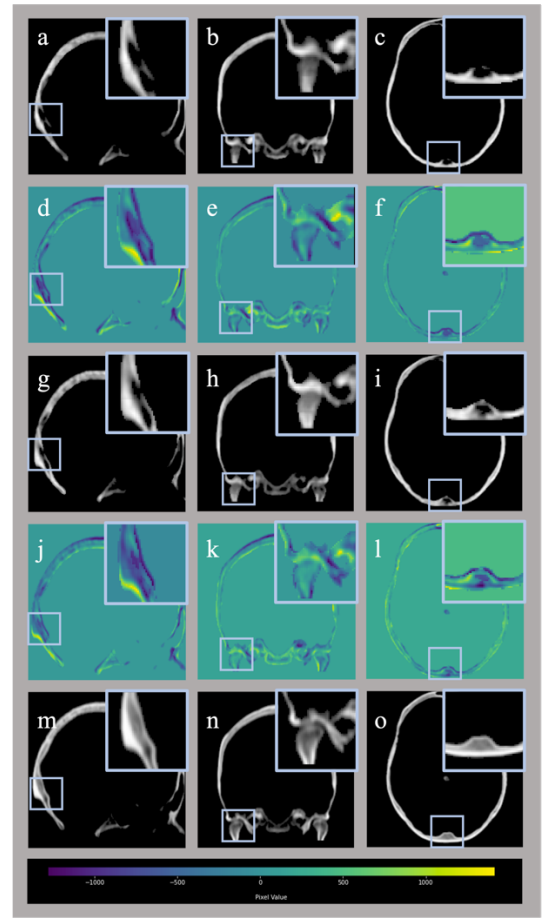

**Fig. 3.** Visualization of the main task results along three anatomical directions. (a-c) Synthetic CT images based on the single T1w modality, presented in the sagittal, coronal, and transverse planes, respectively. (d-f) Residual maps comparing synthetic CT images based on the T1w modality with real CT images. (g-i) Synthetic CT images based on the T1w+FLAIR dual modality in the sagittal, coronal, and transverse planes, respectively. (j-l) Residual maps between the synthetic CT images based on T1w+FLAIR and real CT images. (m-o) Real CT images in the corresponding planes.

To comprehensively evaluate the performance of model in handling sub-tasks and the main task, we assessed the segmentation task from a geometric perspective using Dice coefficient and Jaccard index. To evaluate continuous pixel value predictions, we applied Pearson correlation, Spearman correlation, SSIM (Structural Similarity Index), and MAE

(Mean Absolute Error) among other indices. We also measured the overall image quality through PSNR (Peak Signal-to-Noise Ratio). **Table 1** summarize the quantitative performance of the two sets of sub-tasks for the test dataset.

**Table1.** Performance on sub-tasks in the test dataset.

| Index | Pixel to pixel prediction | | | Segmentation | |
|---|---|---|---|---|---|
| | SSIM | Pearson | Spearman | Dice | Jaccard |
| T1w | | | | | |
| Sub1 | 0.921 | 0.901 | 0.841 | 0.911 | 0.836 |
| Sub2 | 0.913 | 0.856 | 0.782 | 0.834 | 0.715 |
| Sub3 | 0.935 | 0.900 | 0.847 | 0.908 | 0.831 |
| Sub4 | 0.924 | 0.895 | 0.823 | 0.868 | 0.766 |
| Avg | 0.923 | 0.888 | 0.823 | 0.880 | 0.787 |
| T1w+FLAIR | | | | | |
| Sub1 | 0.948 | 0.943 | 0.902 | 0.928 | 0.865 |
| Sub2 | 0.932 | 0.879 | 0.830 | 0.879 | 0.783 |
| Sub3 | 0.953 | 0.933 | 0.886 | 0.912 | 0.838 |
| Sub4 | 0.945 | 0.924 | 0.872 | 0.889 | 0.800 |
| Avg | 0.944 | 0.920 | 0.873 | 0.902 | 0.821 |
| p-value | <0.01 | <0.01 | <0.01 | 0.08 | 0.07 |

\* SSIM = Structural similarity index, Pearson = Pearson's correlation, Spearman = Spearman's rank correlation, Dice = Sørensen–Dice coefficient, Jaccard = Jaccard similarity coefficient.

The Transformer U-Net trained on both T1w and T1w+FLAIR modalities consistently outperformed the single-modality T1-trained model across all four test subjects. To assess whether this performance improvement is statistically significant, we conducted paired t-tests. For pixel prediction sub-tasks, all p-values are below 0.01, both for the structure similarity and correlation indices. However, the improvement in morphological evaluation for the segmentation task only showed a trend (Dice and Jaccard coefficients <0.1).

**Table2.** Performance on the main task in the test dataset.

| Index | SSIM | Pearson | Spearman | PSNR | MAE skull (HU) |
|---|---|---|---|---|---|
| T1w | | | | | |
| Sub1 | 0.946 | 0.920 | 0.894 | 24.344 | 297.839 |
| Sub2 | 0.936 | 0.877 | 0.831 | 24.317 | 307.392 |
| Sub3 | 0.955 | 0.921 | 0.894 | 27.226 | 240.872 |
| Sub4 | 0.942 | 0.907 | 0.862 | 24.845 | 313.880 |
| Avg | 0.945 | 0.906 | 0.871 | 25.183 | 289.996 |
| T1w+FLAIR | | | | | |
| Sub1 | 0.956 | 0.949 | 0.916 | 25.248 | 272.004 |
| Sub2 | 0.941 | 0.885 | 0.854 | 24.720 | 298.329 |
| Sub3 | 0.959 | 0.936 | 0.901 | 27.404 | 229.618 |
| Sub4 | 0.950 | 0.923 | 0.882 | 25.299 | 278.186 |
| Avg | 0.951 | 0.923 | 0.888 | 25.668 | 269.534 |
| p-value | <0.05 | <0.05 | <0.05 | <0.05 | <0.05 |

\* PSNR = Peark Signal-to-Noise-Ratio, MAE = Mean Average Error, HU = Hounsfield Unit.

**Table 2** further compares the results for the synthetic CTs generated by the dual-pipeline combined approach. The dual-modality model's performance remains superior to single-modality T1 in all performance metrices(p<0.05). Notably, MAE is evaluated on images rescaled to match CT ground truth Hounsfield Units (HU), and skull MAE is calculated within the actual CT skull region. This signifies a holistic evaluation of the combined segmentation and prediction effects.

## 4. DISCUSSION

Our framework has demonstrated strong performance in both segmentation and prediction tasks. The evaluation metrics reflected the superiority of our multi-task processing framework in handling the MRI-to-CT modality conversion task. The models demonstrated a substantial improvement in performance when combining pixel prediction and mask segmentation with the generation of synthetic CT, as opposed to the individual prediction of CT, as illustrated in Table 1. This shows the significance of segmentation tasks alongside the modality conversion. It is essential to emphasize that all our evaluations were conducted globally in 3D data, as to not rely on biased comparisons of 2D slices. The embedding of 3D patch position extraction alleviates the need for manual padding or cropping of the original input images, eases the computational burden of large-scale 3D images, accelerates model learning, and allows the model to focus on structural details. Furthermore, it enables us to achieve promising results on limited datasets without the need for additional data augmentation, in contrast to previous related studies[14, 22]. However, it is worth noting that the choice of patch size can significantly impact the final training outcomes of the model. In our experiments, we found that selecting patch sizes that are too small may lead to an insufficient grasp of the overall image structure. Additionally, potential "border effects" in image reconstruction based on patches can be mitigated by reducing the step size without affecting the overall assessment.

Through experimental comparisons, we observed that using dual-modality inputs, T1w and FLAIR, significantly improved the model's performance, especially in terms of pixel value prediction and image correlation enhancement, compared to its performance with single T1w modality. Nevertheless, this improvement was not statistically significant in the isolated segmentation task, possibly due to the relatively small size of our test dataset, despite our progress in augmenting the training dataset. Overall, this might be attributed to the different modalities providing pixel information and additional image details from distinct perspectives. However, for structural delineation, which is particularly sensitive to Dice and Jaccard metrics, the similarity of the morphological information could be extracted from two modalities. It is worth nothing that a single metric may not fully capture the image enhancement effect. Additionally, the multi-modal model demands higher training costs and more complex data acquisition steps. If the improvement in model performance is not substantially evident when using additional information, then opting for the single T1 modality might be the preferable choice.

## 5. Conclusion

In this study, we have introduced a multi-task framework for processing multimodal images, encompassing image segmentation, prediction, 3D patch extraction, and image sequential reconstruction. We have explored the advantages of using multi-modal MRI data, over single sequence for the task of synthesizing CT images, utilizing a public human brain dataset. Our findings reveal that the dual-modal T1w+FLAIR images could contribute richer detail, particularly in terms of pixel-level prediction such as Pearson's correlation and MAE, etc. However, the improvements in morphological aspects, such as image segmentation, are relatively modest.

## 6. Compliance with Ethical Standards

The data used in this study are publicly accessible human brain datasets, which have been shared in accordance with the ethical guidelines and regulations of the respective data providers and repositories. As these datasets are publicly available and de-identified, no additional ethical approval was required for their utilization in this research.

## 7. Acknowledgments

The author has no conflict of interest to disclose.